\documentclass[12pt]{iopart}
\usepackage{graphicx}

\begin{document}
\title[Neutrino fluxes from CNO cycle]{Neutrino fluxes from CNO cycle
in the Sun in the non stationary case with mixing.}

\author{A.Kopylov, V.Petukhov}

\address{Institute of Nuclear Research of Russian Academy of Sciences \\
117312 Moscow, Prospect of 60th Anniversary of October Revolution
7A, Russia}

\ead{beril@al20.inr.troitsk.ru}

\begin{abstract}
The computational analyses is presented of the non stationary case
with mixing of the solar model when the neutrino flux $F_{13}$ from the decay of $^{13}N$ is
higher than a standard solar model predicts.
\end{abstract}
\noindent{\it neutrino experiments, solar abundance, solar interior}

\maketitle

\section{Introduction}
Recently it has been argued \cite{1}-- \cite{3} that to achieve
agreement of the solar surface heavy element abundance with the
results of helioseismology \cite{4, 5} probably one needs to
introduce some non stationary process in the solar model. In the
paper \cite{6} we addressed the question how the mixing in the Sun
can change the fluxes of neutrinos generated in the CNO cycle. The
point which drew our attention was that there is a large difference
of abundance of carbon in the central and peripheral layers due to
burning of carbon in thermonuclear reactions in the center of the
Sun. This makes possible to obtain a relatively large increase of
the flux $F_{13}$ from the decay of $^{13}$N even for a very mild
mixing between the central and peripheral layers of the Sun while
all other neutrino fluxes and other observable are not practically
changed. Looking for the effect of mixing we have omitted
coordinates in our consideration neglecting the specific character
of the mixing, paying attention only on the evolutionary part of
equations.

{\footnotesize
\begin{equation}
\label{eq1} \left\{ {\begin{array}{l}
 dX(^{12}C)/dt=-\lambda (^{12}C) X(^{12}C)+\frac{12}{15} \lambda (^{15}N) X(^{15}N)+\\\qquad+k
(X_0 (^{12}C)-X(^{12}C)) \\
 dX(^{13}C)/dt=-\lambda (^{13}C) X(^{13}C)+\frac{13}{12} \lambda (^{12}C) X(^{12}C)-k X(^{13}C) \\
 dX(^{14}N)/dt=-\lambda (^{14}N) X(^{14}N)+\frac{14}{13} \lambda (^{13}C) X(^{13}C)+\\\qquad+k
(X_0 (^{14}N)-X(^{14}N)) \\
 dX(^{15}N)/dt=-\lambda (^{15}N) X(^{15}N)+\frac{15}{14} \lambda (^{14}N) X(^{14}N)-k X(^{15}N) \\
 \end{array}} \right.
\end{equation}

}
\noindent where $X(^{A}Z)$ and $X_0(^{A}Z)$ -- the abundance of isotope $^{A}Z$ in the central 
and peripheral regions, $\lambda (^{A}Z)=\rho N_A X(^1H)<^AZp>$ -- the ($p+^AZ$) reaction
nuclear rate \cite{7,8}, $k$ -- the mass
transport coefficient. Here the mixing is understood as a
spherically asymmetric process when the masses coming to the core of
the Sun and leaving the core go by different paths. So this differs
from gravitational settling and diffusion and in this sense is a 3D
extension of the standard solar model which is known to be a 1D
model. But we followed a simplified approach. We used a weak mixing
(which means that it involves very weak mass transport) which
proceeds fast in comparison with the rates of thermonuclear
reactions in the Sun. This enabled us to divide the internal zone
(0.6 solar radius) of the Sun on spherical layers (onion like
structure) isolated one from another. The set (\ref{eq1}) has been
solved for each spherical layer with the initial conditions of the
standard solar model at the moment when the mixing has been
initiated. Of course the results obtained in this way may serve only
as guidance. Our aim was to get the averaged values and to see how
they differ from the ones given by the standard solar model. Looking
for the solutions of this set of equations we were particularly
cautious not to go beyond the limits for the parameters of the solar
model fixed by helioseismology. This remarkable selectivity of the
mixing to the neutrino fluxes ($F_{13}$ is increased while $F_{15}$ 
from the decay of $^{15}O$ is almost constant) constitutes the clear signature by which the
future experiments may show unambiguously that there is some non
stationary process in the Sun and, consequently, in other stars as
well, provided that uncertainties of the predicted values are less
than the expected changes of the flux. By the present time the main
contribution to these uncertainties gives the one for the cross
section of the thermonuclear reactions which is around 10{\%}
\cite{9}. The parameters used by the standard solar model (the solar
temperature et al) have a lesser impact. It has been shown in
\cite{10} that there is a correlation of $F_{13}$ and $F_{15}$ with
$F_8$ so that the ratio of $F_{13}$ ($F_{15}$) to $F_8$ has less
uncertainty than the fluxes themselves. The total flux of boron
neutrinos has been measured precisely in the SNO experiment
\cite{11}; this makes possible to reduce the contribution of this
uncertainty. Another important issue is the intensity of the mixing,
i.e. the exact value of $k$. In the paper \cite{6} the results were
obtained for the case with a certain value of a mass transport
coefficient ($k = 10^{-10} yr^{-1}$) and for a certain duration 
$T_{mix}$ of the mixing on the final stage of the solar evolution.
We were tracing how the various parameters of the solar model have
been changed with the duration $T_{mix}$ and compared the found
changes with the limits established by helioseismology. It was shown
that mixing with $k = 10^{-10} yr^{-1}$ and the duration 
$T_{mix}=5\cdot 10^8$ yr is within these limits and this can be used
as a partial solution to find the general one. It means that a mass
0.05 of the internal zone of the Sun can be transported to and from
the central region of the Sun during the final phase of the solar evolution of
the duration $T_{mix}=5\cdot 10^8$ yr. This indicates the intensity
of the mixing process which does not contradict to present data of
helioseismology.

\section{Detailed investigation of the case of mixing}

Here we investigate this case in more details varying freely $k$ and
$T_{mix}$. On Fig. 1 (a-d) the contours are shown for the standard
relative uncertainties 1, 2 and 3 $\delta $ (which correspond to the
uncertainty 0.5{\%}, 1.0{\%} and 1.5{\%} for the mean molecular
weight in the core of the Sun) and for the abundance of hydrogen,
helium, carbon and nitrogen on the surface of the Sun. Here $\delta
$ is a relative uncertainty which corresponds to a standard
deviation $\sigma $. 

\begin{figure}[!ht]
\centering
\includegraphics[width=3in]{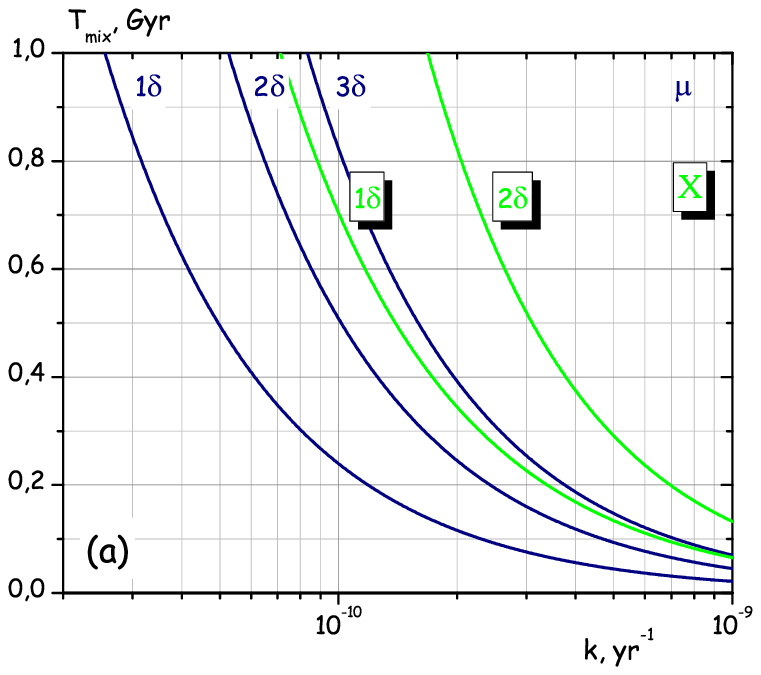}
\includegraphics[width=3in]{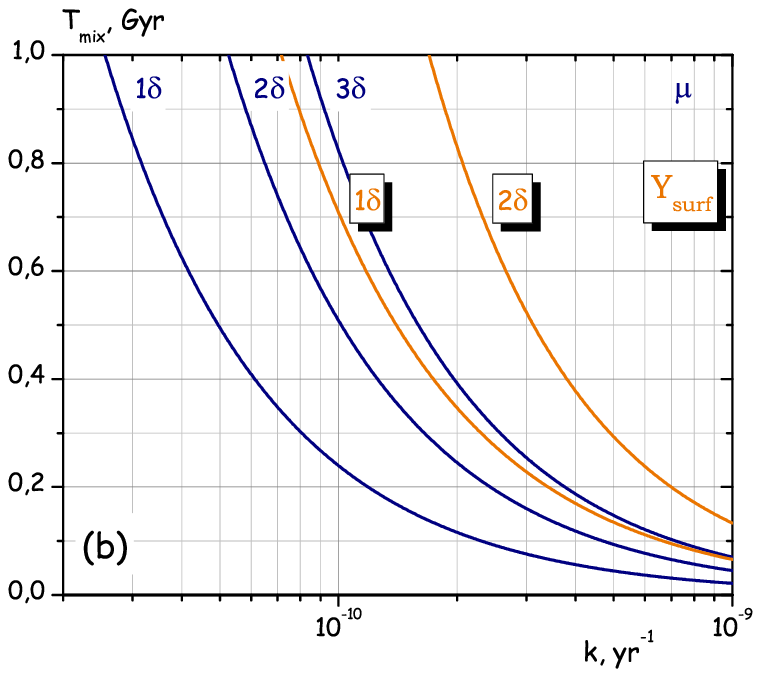}
\includegraphics[width=3in]{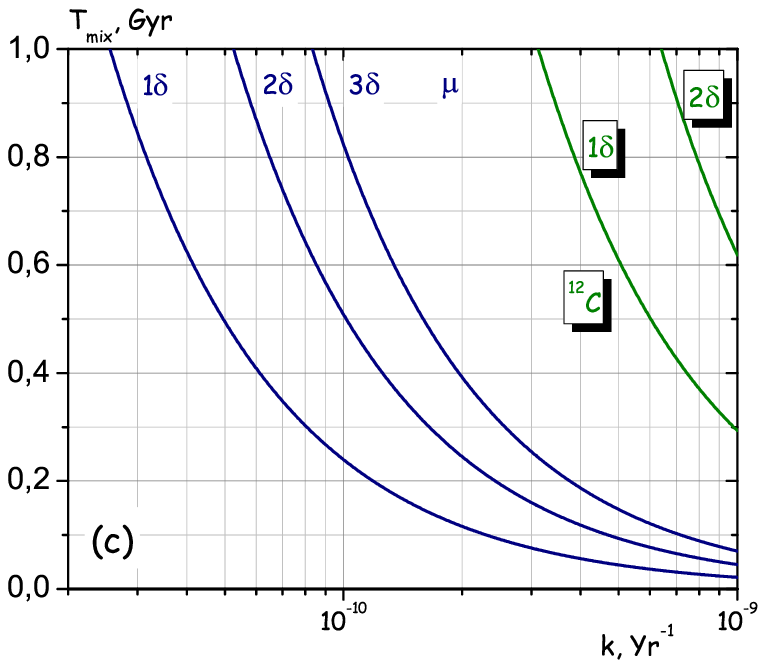}
\includegraphics[width=3in]{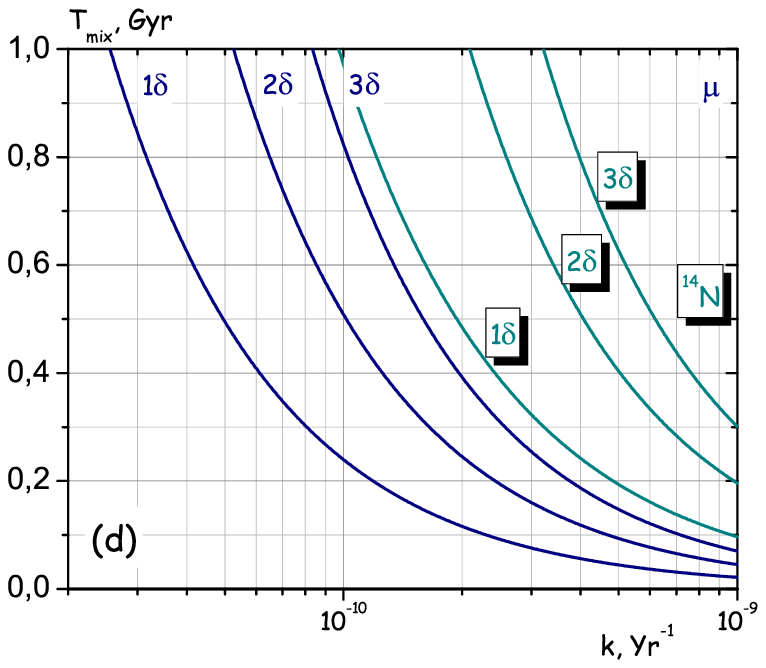}
\caption{The plots for $\mu$ in the core of the Sun and the
abundance of $^1H$ (a), $^4He$ (b), $^{12}C$ (c) and $^{14}N$ (d) on
the surface of the Sun.}
\end{figure}

One can see from this figure that the most
stringent limit comes from the mean molecular weight $\mu$ in the
core of the Sun and is only mildly affected by the abundance of
hydrogen, helium and nitrogen on the surface of the Sun. This result
is a good illustration of the power of helioseismology: what concerns the
hypothetical scenario the structure of the center of the Sun turns out to be more crucial than what is
observed on the surface of the Sun. Figure 2 shows the plots for the increase of
the flux of $F_{13}$ (a) and decrease of $F_{15}$ (b) relative to
the ones given by a standard solar model.

\begin{figure}[!ht]
\centering
\includegraphics[width=3in]{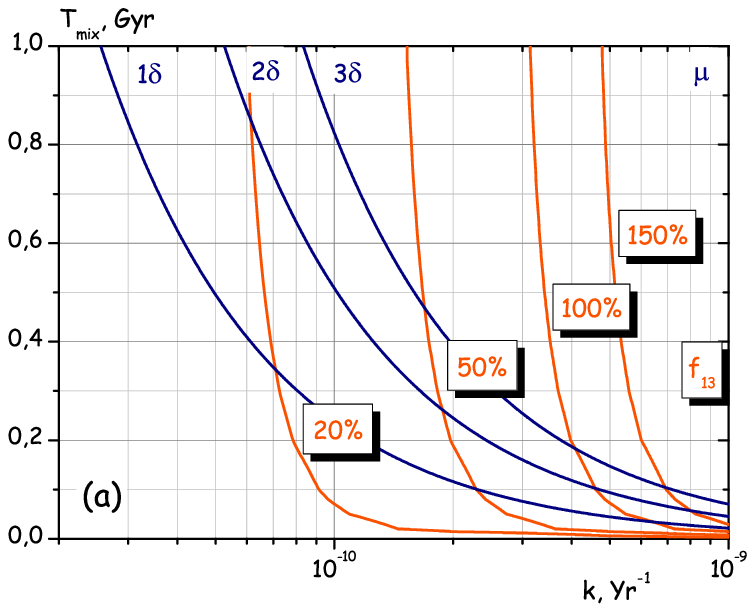}
\includegraphics[width=3in]{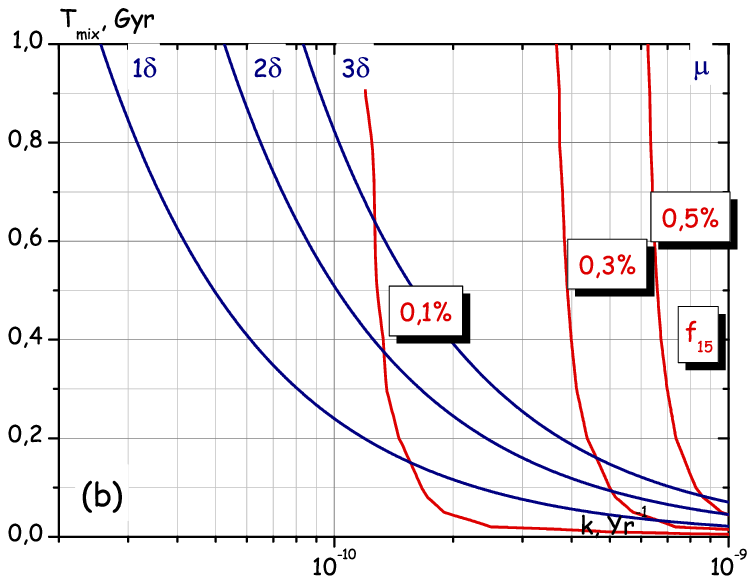}
\caption{The plots for $\mu $ in the core of the Sun and the ones
for the increase of $F_{13}$ (a) and decrease of $F_{15}$ (b).}
\end{figure}

One can see that substantial increase of $F_{13}$ can be gained
by mixing while $F_{15}$ does not experience any noticeable change.
The mixing can change also the distribution of $^3He$ across the
radius of the Sun and, consequently, the flux of $^7Be$ neutrinos.
The new set of differential equations analogous to (\ref{eq1}) can
be written to find these changes.

{\footnotesize
\begin{equation}
\label{eq2}\left\{ {\begin{array}{l}
 dX(^1H)/dt=-X^2(^1H)\frac{3}{2}(\alpha_{11}+\alpha_{11}')
+X^2(^3He)\frac{1}{9}\alpha_{33}-\\\qquad-X(^3He)X(^4He)\frac{1}{12}\alpha_{34}+k(X_0(^1H)-X(^1H)) \\
 dX(^3He)/dt=-X^2(^3He)\frac{1}{3}\alpha_{33}-X(^3He)X(^4He)\frac{1}{4}\alpha_{34}
+\\\qquad+X^2(^1H)\frac{3}{2}(\alpha_{11}+\alpha_{11}')+k(X_0(^3He)-X(^3He)) \\
 dX(^4He)/dt=X^2(^3He)\frac{2}{9}\alpha_{33}+X(^3He)X(^4He)\frac{1}{3}\alpha_{34}
+\\\qquad+k(X_0(^4He)-X(^4He)) \\
 \end{array}} \right.
\end{equation}

}
\noindent where $\alpha_{ij}=\rho N_A <A_iA_j>$ is the ($A_i+A_j$) reaction
nuclear rate \cite{7,8}. The flux $F_7$ of $^7Be$ neutrinos on the Earth can be written as

\begin{equation}
\label{eq3} \Phi_{nu}(^7Be)=\frac{R_{\odot}^3 \cdot N_A}{3\cdot
4}\frac{1}{L^2}\int\limits_0^1 {r^2\rho(r)X_{^3He}(r,t)X_{^4He}
(r,t)} \alpha_{34}(r)dr
\end{equation}

\noindent where $R_{\odot}$ -- solar radius, $L$ -- Sun--Earth distance.

One can see from Fig.3 that the mixing compatible with the results
of helioseismology can change the flux of $^7Be$-neutrinos by less
than 2{\%}.

\begin{figure}[!ht]
\centering
\includegraphics[width=4in]{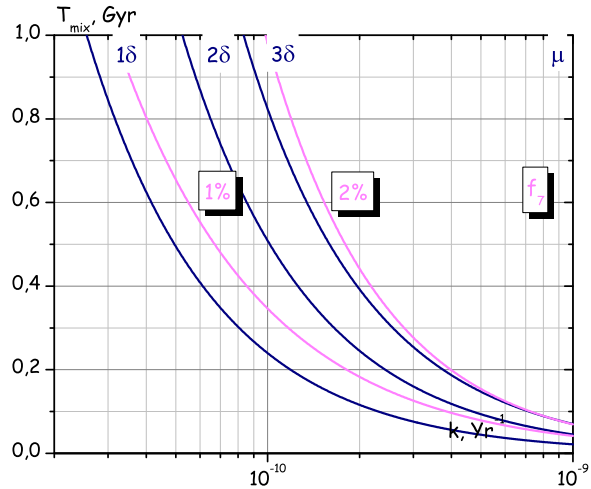}
\caption{The plots for $\mu $ in the core of the Sun and the ones
for the increase of $F_{7}$.}
\end{figure}

\section{Conclusion}

One can see from this analysis that a small freedom is left now
by helioseismology for a mixing of a solar matter. For periods
comparable with the age of the Sun the mixing can be realized only
with $k \le 10^{-10} yr^{-1}$. For greater values of $k$ the mixing
can exist only as the relatively short periods in the Sun's history.
However, the remarkable signature of the mixing is the increase of
the flux $F_{13}$ while all other neutrino fluxes are practically
left unchanged. This conclusion is of general character in a sense
that it does not depend upon the specific kind of mixing in
coordinates of the internal structure of the Sun. From the
experimental point of view it is very interesting to measure
precisely the fluxes $F_{13}$ and $F_{15}$. It has a discovering
potential on the non stationary process in the Sun and other stars
as well.

\section{Acknowledgements}

This work was supported by RFBR grant {\#}07-02-00136A and by a grant of
Leading Scientific Schools of Russia {\#}959.2008.2.

\end{document}